\documentclass[prl,showpacs,twocolumn,amsmath,amssymb,superscriptaddress,preprintnumbers,aps]{revtex4-1}

\usepackage{amssymb}
\usepackage{graphicx}

\begin{document}

\title{M\"obius transformations and electronic transport properties of large disorderless networks}

\author{Yu Jiang} 
%
\author{M. Mart\'{\i}nez-Mares}
%
\author{E. Casta\~no}
\affiliation{Departamento de F\'{\i}sica, Universidad Aut\'onoma
Metropolitana-Iztapalapa, A. P. 55-534, 09340 M\'exico D. F., Mexico}
\author{A. Robledo}
\affiliation{Instituto de F\'{\i}sica, Universidad Nacional Aut\'onoma 
de M\'exico, A. P. 20-364, 01000 M\'exico D. F., Mexico}

\begin{abstract}
We show that the key transport states, insulating and conducting, of  
large regular networks of scatterers can be described generically by  
negative and zero Lyapunov exponents, respectively, of M\"obius maps  
that relate the scattering matrix of systems with successive sizes.  
The conductive phase is represented by weakly chaotic attractors that  
have been linked with anomalous transport and ergodicity breaking. Our  
conclusions, verified for serial as well as parallel stub and ring  
structures, reveal that mesoscopic behavior results from a drastic  
reduction of degrees of freedom.
\end{abstract}


\maketitle


We describe a remarkable relationship between electronic transport in  
regular arrays of scatterers and a special class of low-dimensional  
nonlinear dynamical systems characterized by weak chaos \cite{Geisel,Korabel}. This  
link between two disciplines throws light into the nature of the  
insulator-conductor transition in condensed matter physics, while the  
currently studied field of weak chaos that exhibits anomalous  
diffusion and ergodicity breaking is provided with a physical  
application \cite{Geisel,Korabel}. Quantum transport  
properties have become of interest due to their fundamental importance  
in the development of nanotechnology; for example, the stability  
of spintronic devices based on quantum networks has been investigated~\cite{Foldi}, quantum interference phenomena such as Aharonov-Bohm oscillations  
in conductance, band formation, and metal-insulator transition in  
disorderless networks, have been studied both experimentally and  
theoretically~\cite{Schopfer,Texier,Singha1994,MMR09}. Quantum networks are also 
important as theoretical models of molecular devices~\cite{Aharony} and mesoscopic systems~\cite{Exner,Kottos}. The latter are experimentally available due to the advancement in microfabrication as well as to availability of auxiliary tools in the  
microwave, acoustic, elastic, and optical domains~\cite{Hul,Maynard,Gutierrez,Ghulinyan}. 

In spite of the importance of understanding the propagation of  
electron waves through large networks of quantum wires with regular or  
disordered structures, the study of the interplay between the  
individual scatterers and their geometric arrangement, that as a whole  
results in complex electronic transport behavior~\cite{Marvel2009}, has been the  
object of less exploration. It is, therefore, pertinent to develop a  
procedure to explicitly appraise the interrelations of these two  
nontrivial facets of the scattering processes. A first step in this  
direction is the determination of the generic features of transport  
due solely to the network structural design. Recent advances~\cite{Marvel2009,Ott2008,Marvel2009b} on the  
study of coupled limit-cycle and chaotic oscillators may serve,  
unsuspectingly, as a mirror of similar simplifying features found here  
for electronic transport properties on large disorderless networks. In the 
study of diffusively-coupled nonlinear oscillators~\cite{Heagy}, geometric  
network structures are set up by coupling matrices, and the properties  
of these matrices determine the dynamical behavior of these systems,  
independently of the details of each individual oscillator. As for  
globally-coupled limit-cycle oscillators, the occurrence of  
low-dimensional nonlinear dynamics in large phase-oscillator systems,  
observed sometime ago, has been explained only recently by the role  
that M\"obius maps play in controlling the dynamics of these systems~\cite{Marvel2009b}. Since fractional linear transformations are invertible, the  
possibility of chaotic behavior would be evidently attributed to the  
time evolution of matrix parameters. 

In this communication we report a general phase transition scenario in electronic transport in networks of scatterers connected either in series or in parallel. Specifically, we show, by using the scattering matrix approach, that the conducting and insulating phases of the mesoscopic systems under consideration can be predicted by the behavior of the \emph{finite-time} Lyapunov exponent of a nonlinear map in the complex plane, of the M\"obius group type, that represents the recursive relation between the scattering matrices of successive size generations of a network structure. If we regard the system generation index $n$ as the number of map iterations, then the dynamical behavior can be used to describe in a quantitative manner the electronic transport properties of the quantum mesoscopic system.

\begin{figure}
\includegraphics[width=4.9cm]{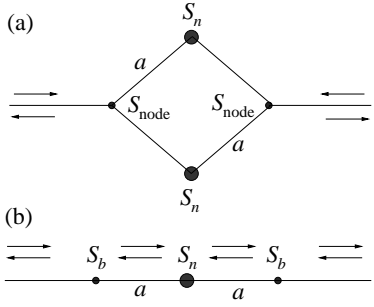}
\caption{Construction of regular mesoscopic networks of scatterers in parallel (a) or in series (b) by means of three-terminal (a) or two-terminal (b) junctions, where $a$ is the lattice constant.}
\label{fig:cayleyd1}
\end{figure}

We consider symmetrical networks constructed by putting together a collection of individual scatterers as building blocks, shown schematically in Fig.~\ref{fig:cayleyd1}. The assembled networks have only two end-points and the single scatterers when connected in series have two terminals, but when connected in parallel have multiple terminals. 
A simple model for three-terminal junctions (connectivity $K=2$) is described by the scattering matrix 
\cite{Buettiker}
\begin{equation}
S_{\rm{node}} = \left[ 
\begin{array}{ccc}
-(\alpha+\beta) & \sqrt{\epsilon} & \sqrt{\epsilon} \\ 
\sqrt{\epsilon} & \alpha & \beta \\
\sqrt{\epsilon} & \beta & \alpha 
\end{array} 
\right],
\end{equation}
where $\epsilon$, $\alpha$, and $\beta$  
are real parameters, related to the transmission and reflection amplitudes of the node:     $0\le\epsilon\le 1/2$, $\alpha=-(1-\sqrt{1-2\epsilon})/2$, and $\beta=(1+\sqrt{1-2\epsilon})/2$. 
When these nodes are repeatedly connected in parallel, keeping one initial terminal free, a Cayley tree \cite{Domb} is formed, and the symmetrical network with two end-points consists of a double Cayley tree joined by two-terminal individual scatterers described by the 
matrix  
\begin{equation}
S_b = \left(
\begin{array}{cc} 
r_b & t'_b \\ t_b & r_b 
\end{array}
\right), 
\end{equation}
which is $2\times 2$, where $r_{b}$ and $t_{b}$ are reflection and transmission amplitudes, respectively.
Then the scattering matrix $S_{n+1}$ of a network of $2(n+1)$ scatterers obtained by doubling the size of a previous generation network of $2n$ scatterers with scattering matrix $S_{n}$ is given by \cite{MMR09}
\begin{equation}
\label{eq:Kparallel}
S_{n+1} = \frac{-1}{\openone-\sqrt{1-2\epsilon}\, e^{2ika}S_n}
\left( \sqrt{1-2\epsilon} \openone - e^{2ika}S_n \right), 
\end{equation}
where $\openone$ is the $2\times 2$ unit matrix and $a$ is the lattice constant.

Clearly, the network with serial connections is a chain. The scattering matrix $S_{n+1}$ of a chain network of $2(n+1)$ scatterers obtained by connecting end-to-end two identical scatterers $S_b$ with scattering matrix $S_{n}$ of $2n$ scatterers is given by
\begin{equation}
\label{eq:series}
S_{n+1} = \frac{1}{\openone - e^{2ika} r_b S_n} 
\left[ r_b \openone - e^{2ika} \left(r_b^2 - t_bt'_b \right)S_n \right].
\end{equation}

For elastic scattering the matrix $S_n$ must be unitary as a result of flux conservation, and has the general form
\begin{equation}
S_n = \left( \begin{array}{cc}
r_n & t'_n\\
t_n & r_n
\end{array} \right), 
\end{equation}
which can be diagonalized through the similarity transformation $S'_n=US_nU^{\dagger}$, where $U$ is the unitary matrix
\begin{equation}
U = \frac{1}{\sqrt{2}} \left( 
\begin{array}{cc}
1 & \sqrt{t'_n/t_n} \\ -\sqrt{t_n/t'_n} & 1 
\end{array} \right),
\end{equation}
and 
\begin{equation}
S'_n = \left( \begin{array}{cc}
e^{i\theta_n} & 0 \\ 0 & e^{i\theta'_n} 
\end{array} \right), 
\end{equation}
where $\theta_n$ and $\theta'_n$ are the eigenphases, which are related to the reflection and transmission amplitudes through $e^{i\theta_n}=r_n+\sqrt{t_nt'_n}$, 
$e^{i\theta'_n}=r_n-\sqrt{t_nt'_n}$, and consequently, the dimensionless conductance (electronic conductance $G$ in units of $2e^2/h$) can be written as
$g_n = \vert t_n\vert^2=\vert t'_n\vert^2= 
\frac{1}{4} | e^{i\theta_n}-e^{\theta'_n} |^2$.

We rewrite the recursive relations (\ref{eq:Kparallel}) and (\ref{eq:series}) as 
one-dimensional maps for the eigenphases: $\theta_{n+1}=f(\theta_n)$ (the map for $\theta'_n$ is identical). These maps are fractional linear transformations of the form
\begin{equation}
\label{eq:Mobius}
z_{n+1} = F(z_n) = \frac{Az_n+B}{Cz_n+D}, 
\end{equation}
where $z_n=e^{i\theta_n}$ and $A$, $B$, $C$ and $D$ are complex numbers: 
$A=e^{2ika}$, $B=-\sqrt{1-2\epsilon}$, $C=-\sqrt{1-2\epsilon}\,e^{2ika}$, and $D=1$ for the double Cayley tree, while for the chain network $A=-(r_b^2-t'_bt_b)e^{2ika}$, $B=r_b$, $C=-r_be^{2ika}$, and $D=1$. The transformation $F(z)$ and its inverse are analytic on the unit circle in the complex plane, and via functional composition define a subgroup (called M\"obius group) that maps one-to-one the unit circle onto itself. The finite $n$ Lyapunov exponent $\lambda_n$ associated to this map is defined by
\begin{equation}
\lambda_n = 
\frac{1}{n}\log 
\left| \frac{df(\theta_n)}{d\theta_0} \right|
\equiv 
\frac{1}{n}\log 
\left| \frac{dz_{n+1}}{d\theta_0} \right|,
\end{equation}
where $\theta_0$ is an initial condition.  

\begin{figure}
\includegraphics[width=4.0cm]{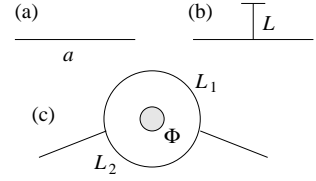}
\caption{
Three different junctions described by an scattering matrix $S_b$: (a) a geometric connection, (b) a stub, and (c) an Aharonov-Bohm ring threaded by a magnetic field.}
\label{fig:casesSb}
\end{figure}

The general form (\ref{eq:Mobius}) is due to the symmetry of the networks together with the uniform distribution of scatterers, and for this reason, in principle, we can generalize the model networks to a class of networks whose growth in size is characterized by successive scattering matrices generated by M\"obius actions. 
In what follows we report numerical results that verify the relationship between the conductance of a network and the finite $n$ Lyapunov exponent of its associated map $f(\theta_n)$, for which the number of iteration time steps is the generation index $n$ that measures the size of the network. To demonstrate the generality of such a relationship we consider different types of junctions. 

(i) \emph{A double Cayley tree with connectivity K}.
The electronic transport properties of this structure has been reported in Ref.  \onlinecite{MMR09} for $K=2$. Also, for this case, we consider the following three different junctions (see Fig.~\ref{fig:casesSb}): (a) A geometric connection with the central scatterer defined by $S_b=\sigma_x e^{ika}$, with $\sigma_x$ a Pauli matrix;  
(b) a stub or quantum gate of length $L$ defined by \cite{Xia}
\begin{equation}
\label{eq:stub}
r_b=-\frac{i}{2\tan kL+ i}, \quad 
t_b=t'_b=\frac{2\tan kL}{2\tan kL+ i};  
\end{equation}
and (c) an Aharonov-Bohm ring threaded by a magnetic field whose scattering matrix elements are given by \cite{Xia}
\begin{eqnarray}
r_b & = & \frac{1}{\Delta}\left[ 
e^{-ik_1\Delta L} + e^{ik_2\Delta L} - 
4\left( e^{-i\Delta kL_1} + e^{i\Delta kL_2} \right) \right. 
\nonumber \\ 
& & \left. \quad + 
3\left( e^{-ik_2L_1-ik_1L_2} + e^{ik_1L_1+ik_2L_2}  \right)
\right] , 
\nonumber  \\ \label{eq:AB1}
t_b & = & \frac{4}{\Delta} \left[
e^{-i\Delta kL_1+ik_2L_2} - e^{-i\Delta kL_1-ik_1L_2} \right. 
\\  & & \quad \left. + 
e^{-ik_1\Delta L+ik_2L_2} - e^{ik_2\Delta L+ik_1L_2}
\right], \nonumber  
\end{eqnarray}
and $t'_b=-r_bt_b/r_b^*$, with $\Delta=e^{-ik_1\Delta L}+e^{ik_2\Delta L}- e^{ik_1L_1+ik_2L_2}+4(e^{-i\Delta kL_1}+e^{i\Delta kL_2})- 
9(e^{-ik_1L_2-ik_2L_1})$, where $L_1$ and $L_2$ are the lengths of the upper and the lower arm  of the ring, and $\Delta L=L_2-L_1$. Here, the wave vectors are given by 
$k_1=k+2\pi\Phi e/hcL$ and $k_2=k-2\pi\Phi e/hcL$, where $L=L_1+L_2$, $\Phi$ is the magnetic flux through the ring, and  $\Delta k=k_2-k_1=-4\pi e\Phi/hcL$.

\begin{figure}
\includegraphics[width=6.4cm]{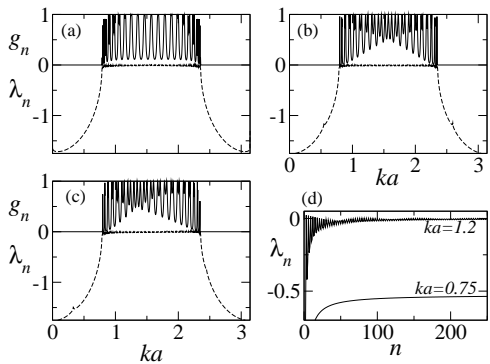}
\caption{
The dimensionless conductance $g_n$ (solid lines, $n=20$) and the finite $n$ Lyapunov exponent $\lambda_n$ (dashed lines, $n=100$) are plotted as a function of $ka$ in a network formed by a double Cayley tree, when the connecting scattering centers are given by (a) a perfect conducting line, (b) a stub, and (c) a ring threaded by a magnetic field ($L=a$, $L_1=L_2=L/2$, $\Phi=0.1hc/e$). (d) Convergence of the finite $n$ Lyapunov exponent $\lambda_n$ to its limiting value $\lambda$. The sign alternations for $ka=1.2$ indicate the intermittent quality of the conductive phase.}
\label{fig:Fig2}
\end{figure}

In Fig.~\ref{fig:Fig2} we show the relationship between the conductance and the finite $n$ Lyapunov exponent for very large $n$. We see that for each case the Lyapunov exponent is negative for small values of $ka$, approaches zero as $ka$ increases, and the network is in an insulating phase with $g_n=0$. The exponent $\lambda_n$ reaches zero at some critical value of $ka$, and at the same point the mesoscopic system undergoes a transition from an insulating to a conducting state. As $ka$ is further increased $g_n>0$ while $\lambda_n=0$ [$|d\theta_{n+1}/d\theta_0|/n$ oscillates but becomes zero for $n\rightarrow\infty$, as shown in Fig.~\ref{fig:Fig2}(d)], until at a second critical value the system returns to the insulating state and $\lambda_n<0$. There is a remarkable, perfect, correspondence between the transport and the dynamical behavior of the nonlinear maps that represent the three different junctions in Fig.~\ref{fig:casesSb}~\cite{MMR09}. 

An extrapolation of the expression for the recursive relation of scattering matrix, or the eigenphase map $f(\theta_n)$, for the general case of arbitrary $K$, indicates that the metal-insulating transitions for $K>2$ can be deduced from the $K=2$ case simply by replacing $\epsilon$ by $2\epsilon/K$. 

\begin{figure}
\includegraphics[width=5.0cm]{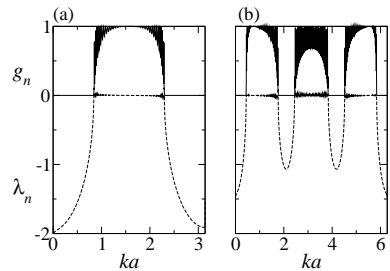}
\caption{
The dimensionless conductance $g_n$ (solid lines, $n=20$) and the finite $n$ Lyapunov exponent $\lambda_n$ (dashed lines, $n=100$) are plotted as a function of $ka$ in a network formed by a chain of serially-connected scatterers of (a) stubs, and (b) rings threaded by a magnetic field.}
\label{fig:Fig3}
\end{figure}

(ii) \emph{A chain of serially-connected scatterers}. 
The chain network when the scatterers are stubs, as in Fig.~\ref{fig:casesSb}(b), has been studied in Ref.~\onlinecite{Singha1994} by using the transmission matrix method. Here we analyze its phase transition properties via the scattering matrix approach, and in this way reveal some common transport properties of mesoscopic systems with differing types of scatterers. Fig.~\ref{fig:Fig3}(a) exhibits the same transition scenarios as observed in the double Cayley tree. A linear network may appear to be a more realistic structure than the double Cayley tree, but the characteristic behavior the Lyapunov exponent appears again to be a precise indicator of the metal-insulating transitions \cite{MMR09}. 
For an array of mesoscopic rings threaded by magnetic flux as in Fig.~\ref{fig:casesSb}(c) the elements of the building block scattering matrix are given by Eqs. (\ref{eq:AB1}). The numerical results shown in Fig.~\ref{fig:Fig3}(b) reveal once more the transition scenarios described previously. The positions where the Lyapunov exponent becomes zero indicate quantitatively the locations of the metal-insulating transtions that occur in the system of serially connected rings.

Clearly, our results provide convincing evidence that low-dimensional dynamical transitions, from a stable fixed-point to weakly chaotic attractors, with negative and vanishing Lyapunov exponents, respectively, are indicators of insulating-conducting transitions occurring in solid-state model systems. 
Interestingly, the vanishing of the ordinary exponent $\lambda$ takes place via intermittency in the sign of the finite $n$ exponent $\lambda_n$, a signature of weak chaos in maps with a tangency feature like ours~\cite{MMR09,Korabel}. 
This constitutes a physical property of the conducting networks, i.e.  
the conductance oscillates when finite size is changed. (A unit increment in $n$ represents at least a doubling of size in the linear chain).
Our development also points out a rare simplification in which there is a large reduction of degrees of freedom: a system composed by many scatterers is described by a low-dimensional map.
This property is reminiscent of the drastic reduction in state variables displayed by large arrays of coupled limit-cycle oscillators, for which their macroscopic time evolution has been shown \cite{Marvel2009b} to be governed by underlying low-dimensional nonlinear maps of the form of Eq. (\ref{eq:Mobius}).
The parallelism between the transport properties of the networks studied here and the dynamical properties of arrays of oscillators can be made more specific by noticing that in both problems the variables of interest are comparable, phase shift of the scattered states and phase time change of coupled oscillators, both determined by the action of the M\"obius group. In the scattering problem we arrive at the basic nonlinear map by first constructing a family of self-similar networks arranged by size, or generation $n$, and then relating the scattering matrices of two consecutive generations. In the case of coupled oscillators the puzzle of the drastic reduction of variables finds a rationale in the identification of the role of the M\"obius group in the temporal evolution of the system. Notice that the particulars of the scattering potentials appear only through the coefficients in Eq. (\ref{eq:Mobius}) while the structure of the network gives the transformation its general form.

\begin{figure}
\makebox{\includegraphics[width=4.6cm]{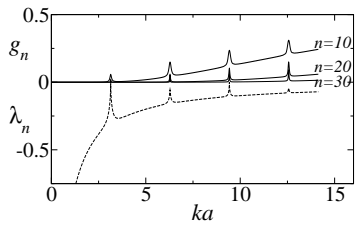}}
\caption{
The dimensionless conductance $g_n$ (solid) and the finite $n$ Lyapunov exponent $\lambda_n$ (dashed, $n=100$) are plotted as a function of $ka$ for a chain of delta wells of imaginary intensity $-iv$ where $v=\hbar^2/2m$.}
\label{fig:Fig4}
\end{figure}

Because the basic recursive relation is a consequence of special  
symmetries and an underlying group-theoretic structure, it is  
interesting to explore the relevance of our results in other symmetric  
networks, including two-dimensional arrays of interconnected  
scatterers. We notice that the correspondence between zero Lyapunov  
exponent and conducting phase does not hold for scattering with  
absorption. In this case the scattering matrix is not unitary and, therefore, the M\"obius group theory does not apply; the corresponding results are shown in Fig.~\ref{fig:Fig4}. As for asymmetrical arrangements of  
assemblies of scatterers it is expected that chaotic or other  
complicated scenarios may be observed, as the recursive relation might  
not be invertible. 


Furthermore, a significant generic finding, as it is independent of the nature of 
the network scatterers, is that the conducting phase is characterized by the 
features of \emph{weak chaos}, that is, ergocidity breaking, infinite invariant measures, and anomalous transport. Experimental realizations are feasible nowdays via the use of optical lattices as models of condensed matter systems~\cite{Courtade}. 
The  effect of  disorder  is  left  for  future  studies. 



\begin{thebibliography}{99}

\bibitem{Geisel}
T. Geisel, S. Thomae, Phys. Rev. Lett. \textbf{52}, 1936 (1984).

\bibitem{Korabel}
N. Korabel, E. Barkai, Phys. Rev. Lett. \textbf{102}, 050601 (2009).

\bibitem{Foldi}
P. F\"oldi, O. K\'alm\'an, and F. M. Peeters, 
Phys. Rev. B \textbf{80}, 125324 (2009).

\bibitem{Schopfer}
F. Schopfer, F. Mallet, D. Mailly, C. Texier, G. Montambaux, C. B\"auerle, and L. Saminadayar, 
Phys. Rev. Lett. \textbf{98}, 026807 (2007).

\bibitem{Texier}
C. Texier, P. Delplace, and G. Montambaux, 
Phys. Rev. B \textbf{80}, 205413 (2009).

\bibitem{Singha1994}
P. Singha Deo and A. M. Jayannavar,  
Phys. Rev. B {\bf 50}, 11629 (1994).

\bibitem{MMR09}
M. Mart\'inez-Mares and A. Robledo,  
Phys. Rev. E {\bf 80}, 045201(R) (2009).

\bibitem{Aharony}
A. Aharony and O. Entin-Wohlman, 
J. Phys. Chem. B \textbf{113}, 3676 (2009).

\bibitem{Exner}
P. Exner, M. Tater, and D. Van\v ek, 
J. Math. Phys. \textbf{42}, 4050 (2001).

\bibitem{Kottos}
T. Kottos and U. Smilansky, 
Phys. Rev. Lett. \textbf{79}, 4794 (1997).

\bibitem{Hul}
O. Hul, O. Tymoshchuk, S. Bauch. P. M. Koch, and L. Sirko, 
J. Phys. A: Math. Gen. \textbf{38}, 10489 (2005).

\bibitem{Maynard}
J. D. Maynard, 
Rev. Mod. Phys. \textbf{73}, 401 (2001).

\bibitem{Gutierrez}
A. Morales, J. Flores, L. Guti\'errez, and R. A. M\'endez-S\'anchez, 
J. Acoust. Soc. Am. \textbf{112}, 1961 (2002). 

\bibitem{Ghulinyan}
M. Ghulinyan, C. J. Oton, Z. Gaburro, L. Pavesi, C. Toninelli, and D. S. Wiersma, 
Phys. Rev. Lett. \textbf{94}, 127401 (2005).

\bibitem{Marvel2009}
S. A. Marvel and S. H. Strogatz,
Chaos {\bf 19}, 013132 (2009).

\bibitem{Ott2008}
E. Ott and T. M. Antonsen 
Chaos \textbf{18}, 037113 (2008).

\bibitem{Marvel2009b}
S. A. Marvel, R. E. Mirollo, and S. H. Strogatz,
Chaos {\bf 19}, 043104 (2009).

\bibitem{Heagy}
J. F. Heagy, T. L. Carroll, and L. M. Pecora, 
Phys. Rev. E \textbf{50}, 1874 (1994).

\bibitem{Buettiker}
M. B\"uttiker, Y. Imry, and M. Ya. Azbel, Phys. Rev. A \textbf{30}, 1982 (1984).

\bibitem{Xia}
J.-B. Xia, 
Phys. Rev. B \textbf{45}, 3593 (1992).

\bibitem{Domb}
C. Domb, in \emph{Phase Transitions and Critical Phenomena} Vol. \textbf{3}, 
edited by C. Domb and M. S. Green (Academic, New York, 1974).

\bibitem{Courtade}
E. Courtade, O. Houde, J.-F. Cl\'ement, P. Verkerk, and D. Hennequin, 
Phys. Rev. A \textbf{74} 031403(R) (2006).


\end{thebibliography}
\end{document}